\documentclass{PoS}

\newcommand{\be}{\begin{equation}}
\newcommand{\ee}{\end{equation}}
\newcommand{\beq}{\begin{eqnarray}}
\newcommand{\eeq}{\end{eqnarry}}

\newcommand{\C}{{\kern+.25em\sf{C}\kern-.45em\sf{{\small{I}}}\kern+.45em\kern-.25em}}
\newcommand{\R}{{\kern+.25em\sf{R}\kern-.78em\sf{I} \kern+.78em\kern-.25em}}

\def\lsi{\raise0.3ex\hbox{$<$\kern-0.75em\raise-1.1ex\hbox{$\sim$}}}
\def\gsi{\raise0.3ex\hbox{$>$\kern-0.75em\raise-1.1ex\hbox{$\sim$}}}

\title{A Non-Perturbative Operator Product Expansion}

\ShortTitle{Non-Perturbative OPE}

\author{\speaker{W.\ Bietenholz}$^{\rm \ a}$, N.\ Cundy$^{\rm \ b}$,
M.\ G\"{o}ckeler$^{\rm \ b}$, R.\ Horsley$^{\rm \ c}$, \newline H.\
Perlt$^{\rm \ d}$, D.\ Pleiter$^{\rm \ e}$, P.E.L.\ Rakow$^{\rm \ f}$,
G.\ Schierholz$^{\rm \ b,g}$, \newline A.\ Schiller$^{\rm \ d}$, 
T.\ Streuer$^{\rm \ b}$ and J.M.\ Zanotti$^{\rm \ c}$ 
\thanks{Work supported by the Deutsche Forschungsgemeinschaft (DFG) 
through project FOR 465 ``Forschergruppe 
Gitter-Hadronen-Ph\"{a}nomenologie'', and through
Sonderforschungsbereich SFB/TR55 ``Hadron Physics from Lattice QCD''.
\ \ \ Preprint DESY 09-166, Edinburgh 2009/14, LTH 846}
\\
\ \\
$^{\rm \ a}$ Insituto de Ciencias Nucleares, Universidad Nacional 
Aut\'{o}noma de M\'{e}xico \\
~~~~A.P.\ 70-543, C.P.\ 04510 Distrito Federal, M\'{e}xico \\
$^{\rm \ b}$ Institut f\"{u}r Theoretische Physik,
Universit\"{a}t Regensburg, 93040 Regensburg, Germany \\
$^{\rm \ c}$ School of Physics, University of Edinburgh, Edinburgh EH9
3JZ, United Kingdom \\
$^{\rm \ d}$ Institut f\"{u}r Theoretische Physik,
Universit\"{a}t Leipzig, 04109 Leipzig, Germany \\
$^{\rm \ e}$ John von Neumann Institut f\"{u}r Computing NIC, \\
~~~~Deutsches Elektron-Synchrotron DESY, 15738 Zeuthen, Germany \\
$^{\rm \ f}$ Theoretical Physics Division, Dept.\ of Mathematical Sciences, University of Liverpool, \\ ~~~~Liverpool, L69 3BX,
United Kingdom \\
$^{\rm \ g}$ Deutsches Elektron-Synchrotron DESY, 22603 Hamburg, Germany\\
\ \\
E-mail: \email{wolbi@nucleares.unam.mx} \\ }

\abstract{Nucleon structure functions can be observed in Deep Inelastic
Scattering experiments, but it is an outstanding challenge to confront
them with fully non-perturbative QCD results. For this purpose
we investigate the product of electromagnetic currents (with large
photon momenta) between quark states (of low momenta). By means of an
Operator Product Expansion the structure function can be decomposed
into matrix elements of local operators, and Wilson coefficients.
For consistency both have to be computed non-perturbatively.
Here we present precision results for a set of Wilson
coefficients. They are evaluated from propagators for numerous
quark momenta on the lattice, where the use of chiral fermions
suppresses undesired operator mixing. This over-determines the
Wilson coefficients, but reliable results can be extracted by
means of a Singular Value Decomposition.}

\FullConference{The XXVII International Symposium on Lattice Field Theory - LAT2009\\
		 July 26-31 2009\\
		 Peking University, Beijing, China}

\begin{document}

\section{Deep Inelastic Scattering}

Historically, Deep Inelastic Scattering gave first evidence that
quarks are in fact physical objects. More generally it provides
insight into the hadron structure functions, and thus into the
distribution of energy and spin among the hadron constituents
(see Ref.\ \cite{Dia} for a recent review).

Here we focus on the {\em nucleon structure functions,} which can be
observed for instance by hard leptonic scattering dominated by 
one-photon exchange, as sketched below. This type of scattering
only involves one quark, hence chirality is conserved.
\begin{figure}[h!]
\begin{center}
\includegraphics[angle=0,width=.4\linewidth]{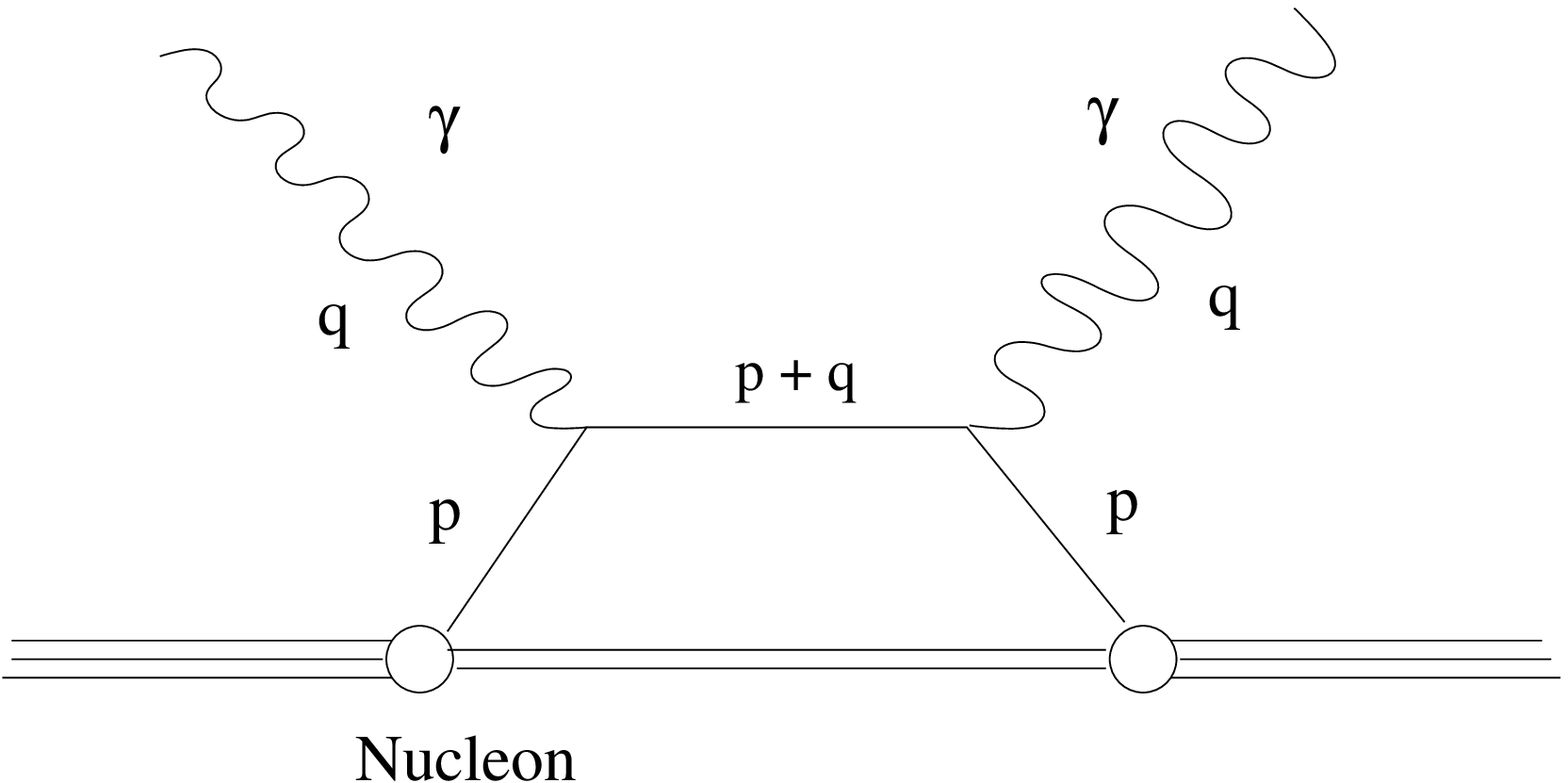} \hspace*{5mm}
\includegraphics[angle=0,width=.4\linewidth]{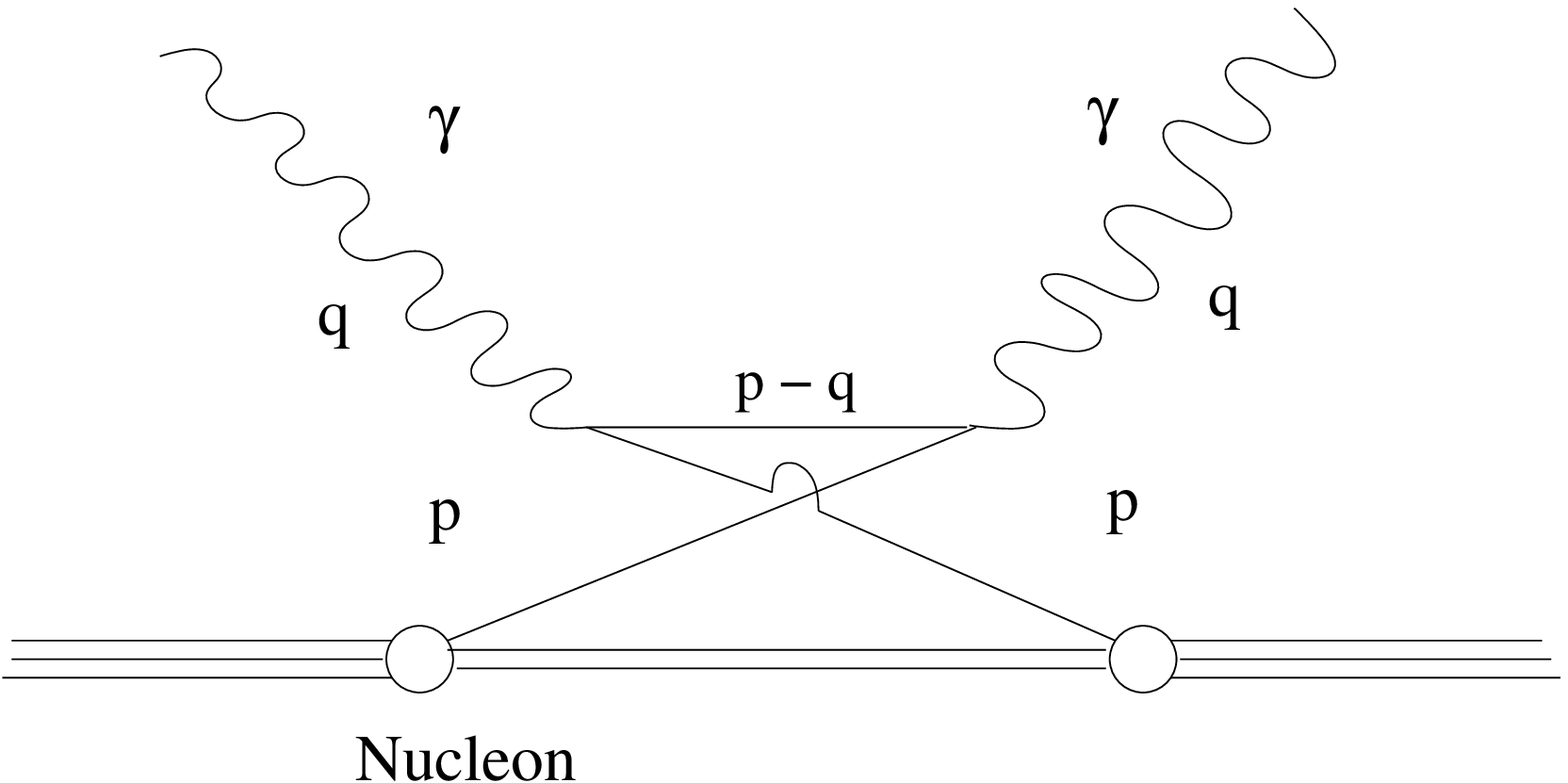}
\end{center}
\vspace*{-3mm}
\end{figure}

\noindent
Despite the high energy in this process, perturbative QCD does not
lead to fully satisfactory results. In particular, power-like IR 
divergences occur, which imply the notorious renormalon ambiguities
\cite{renormalon}. A consistent evaluation of the nucleon structure
function from first principles of QCD has to be fully non-perturbative.
Hence it is a challenge for lattice studies, and the goal of our
project.

\section{Nucleon structure functions on the lattice}

We start from the general ansatz for moments of
a lattice nucleon structure function,
\be \label{nustrufu}                           
{\cal M}(q^{2}) = c^{(2)} (aq) A_{2}(a) + 
\frac{c^{(4)}(aq)}{q^{2}} A_{4}(a) + \ \dots \ \ 
\{ {\rm higher~twists} \} \ ,
\ee
where $a$ is the lattice spacing, $q$ is the photon momentum being 
transferred, $c^{(n)}$ are Wilson coefficients and $A_{n}$ are matrix 
elements 
(their Lorentz structure is factored out). 
Traditionally the latter are computed by lattice simulations, 
whereas the coefficients $c^{(n)}$ are evaluated perturbatively, 
since they only depend on the large photon momentum. 
However, consistency strictly requires the non-perturbative evaluation
of both factors \cite{MaSa}. Here we present precise numerical
results which are relevant for $c^{(2)}$. Further steps in this
project, leading to ${\cal M}(q^{2})$, are reported in 
Refs.\ \cite{Paul,Thomas}.

\section{Lattice technicalities}

We used configurations that were generated quenched
with the L\"{u}scher-Weisz gauge action on a lattice
of size $24^{3} \times 48$ at $\beta = 8.45$.
Based on $r_{0} = 0.5 ~{\rm fm}$, the lattice spacing has been 
determined as $a = 0.095(1)~{\rm fm}$ \cite{laspa}.\footnote{The
error will be ignored in the following. An alternative method based
on $F_{\pi}$ yielded $a \simeq 0.105~{\rm fm}$ \cite{MG}.} 
We fixed the lattice Landau gauge, which is essential 
for obtaining finite values for the matrix elements; moreover
gauge fixing is helpful to reduce the statistical noise.

Our analysis involves two flavours of degenerate valence quarks, 
which are represented by overlap fermions (we apply an overlap 
Dirac operator constructed with a Wilson kernel and a negative
mass shift of $\rho =1.4$). The use of chiral quarks suppresses
$O(a)$ lattice artifacts, as well as undesired operator mixing.
The latter virtue is particularly important in this project;
for instance it is essential to disentangle the contributions
of higher twist, cf.\ eq.\ (\ref{nustrufu}).

Our study includes two bare masses for the
degenerate quark flavours,
\begin{displaymath}
m_{q} = 0.011 \simeq 29 ~{\rm MeV} \ \to \ m_{\pi} \approx 280 ~{\rm MeV}
\quad {\rm and} \quad
m_{q} = 0.028 \simeq 73 ~{\rm MeV} \ \to \ m_{\pi} \approx 440 ~{\rm MeV} \ .
\end{displaymath}

\section{OPE on the lattice}

We use standard lattice electromagnetic currents 
$J_{\mu}$ (it would be computationally expensive to work with currents, 
which are conserved in the framework of overlap quarks). Thus one 
renormalisation constant ($Z_{V}$) will be needed \cite{Paul,Thomas}.
The Operator Product Expansion (OPE) decomposes the product of
two such currents between quark states as follows,
\be  \label{OPEeq}
W_{\mu \nu} \simeq  
\langle \psi (p) | J_{\mu} (q) J_{\nu}^{\dagger} (q)
| \psi (p) \rangle
\overbrace{=}^{\rm OPE} \sum_{m,i,n}             
C_{\mu \nu ,i, \mu_{1} \dots \mu_{n}}^{(m)} (q)
\ \langle \psi (p) | {\cal O}^{(m)}_{i, \mu_{1}
\dots \mu_{n}} | \psi (p) \rangle
\ee
$C^{(m)}$ : Wilson coefficients, independent of the target and therefore 
of the quark momentum $p$ \\
${\cal O}^{(m)}$ : {\em local} operators, relevant to describe the nucleon
structure \\
$\mu_{j}$ : momentum component $p_{\mu_{j}}$ appears in the operator \\
$i = 1 \dots 16$ : Clifford index \, , \
$m$ : index for operators with the same symmetries.

A truncation of the OPE in eq.\ (\ref{OPEeq}), and small lattice
artifacts, require the {\em scale separation}
\be  \label{scasep}
p^{2} \ll q^{2} \ll ( \pi /a )^{2} \ .
\ee
Assuming this separation to hold,
we consider quark bilinears up to $O(|p|^{3})$,
\be
\bar \psi \Gamma \psi \ , \quad \bar \psi \Gamma D_{\mu_{1}} \psi \ , 
\quad \bar \psi \Gamma D_{\mu_{1}} D_{\mu_{2}} \psi \ ,
\quad \bar \psi \Gamma D_{\mu_{1}} D_{\mu_{2}} D_{\mu_{3}} \psi \ .
\ee
The symbol $\Gamma$ captures the full Clifford structure,
hence this set includes a frightening number of
$16 \cdot \sum_{d=0}^{3} 4^{d} = 1360$ operators.
However, we choose specific photon
momenta of the diagonal form $q \propto (1,1,1,1)$, which implies a 
high level of symmetry. To be explicit, we consider 
three photon momenta,
\begin{displaymath}
a q_{\mu}^{(1)} = \frac{\pi}{6} \ , \ 
|q^{(1)}| \simeq 2.2~{\rm GeV} \ \ , \quad
a q_{\mu}^{(2)} = \frac{\pi}{4} \ , \ 
|q^{(2)}| \simeq 3.3~{\rm GeV} \ \ , \quad
a q_{\mu}^{(3)} = \frac{\pi}{3} \ , \ 
|q^{(3)}| \simeq 4.4~{\rm GeV} \ .
\end{displaymath}
For $q^{(2)}$ and $q^{(3)}$ we implement standard boundary conditions
(b.c.), but $q_{\mu}^{(1)}$ is applied along with twisted 
b.c. for the quark fields: in addition to the Euclidean time direction,
also two of the spatial directions are antiperiodic. This gives
access to smaller $p$-momenta, which are needed in view of 
condition (\ref{scasep}), since $|q^{(1)}|$ is not that large.

Thanks to our diagonal choice of $q$, the set of operators 
reduces to only {\em $67$ equivalence classes} \cite{Lat0708}. We 
classify the corresponding Wilson coefficients according to the 
number of derivatives in the operators that they multiply: \\
$C_{1}$ : no derivative, multiplies $\bar \psi 1 \!\! 1 \psi$
\hspace*{9mm}
$C_{2} \dots C_{6}$ : one derivative, 
Bjorken scaling $\propto 1/q^{2}$ \\
$C_{7} \dots C_{16}$ : two derivatives \hspace*{2cm}
$C_{17} \dots C_{67}$ : three derivatives, 
Bjorken scaling $\propto 1/(q^{2})^{2}$. \\
The coefficients of operators with an even number of derivatives 
vanish at $m_{q}=0$ due to chiral symmetry. 
\newpage

In each case, our evaluation of $C_{1} \dots C_{67}$
involves numerous quark momenta $p_{1}, \dots , p_{M}$,
see Table \ref{tab}.
\begin{table}[h!]
\vspace*{-4mm}
\centering
\begin{tabular}{c||c|c|c|c|c|c|}
   & \multicolumn{2}{|c|}{$q^{(1)}$} & \multicolumn{2}{|c|}{$q^{(2)}$} 
& \multicolumn{2}{|c|}{$q^{(3)}$} \\
 & $M$ & $SV_{\rm opt}$ & $M$ & $SV_{\rm opt}$ & $M$ & $SV_{\rm opt}$ \\ 
\hline
$m_{q} = 0.011$ & $15$ & $14$ 
& $31$ & $10$ & $31$ & $10$ \\
\hline
$m_{q} = 0.028$ & $15$ & $12$ 
& $32$ & ~$8$ & $31$ & ~$8$ \\
\hline
\end{tabular}
\caption{The number $M$ of quark momenta, and $SV_{\rm opt}$ of
Singular Values (see below) that we used for the
determination of the Wilson coefficients in each case,
{\it i.e.}\ for each quark mass and photon momentum.}
\label{tab} 
\end{table}
\vspace*{-3mm}

For $q^{(1)}$, {\it i.e.}\ with twisted b.c., there are less
$p$-momenta with small $p^{2}$, hence less $p$-momenta are needed
for converging results.
Thus we measure $W_{\mu \nu}$ --- given in eq.\ (\ref{OPEeq}) ---
off-shell 
for $M= 15 \dots 32$ quark momentum sources
to determine the Wilson coefficients $C_{1} \dots C_{67}$.
Schematically they are given as
(the elements $W^{(p_{i})}$ and ${\cal O}_{k}^{(p_{i})}$ 
are $4 \times 4$ matrices capturing the spin components)
\be  \label{fixC}
\left( \begin{array}{c}
W^{(p_{1})} \\ . \\ . \\ . \\  W^{(p_{M})}
\end{array} \right) = \left( \begin{array}{ccccc}
{\cal O}_{1}^{(p_{1})} & . & . & . & {\cal O}_{67}^{(p_{1})} \\
 . &  . & . & . & . \\
 . &  . & . & . & . \\
 . &  . & . & . & . \\
{\cal O}_{1}^{(p_{M})} & . & . & . & {\cal O}_{67}^{(p_{M})}
\end{array} \right) \
\left( \begin{array}{c}
C_{1} \\ . \\ . \\ C_{67} \end{array} \right)
\ee

\begin{figure}[h!]
\vspace*{-2mm}
\begin{center}
\includegraphics[angle=270,width=.49\linewidth]{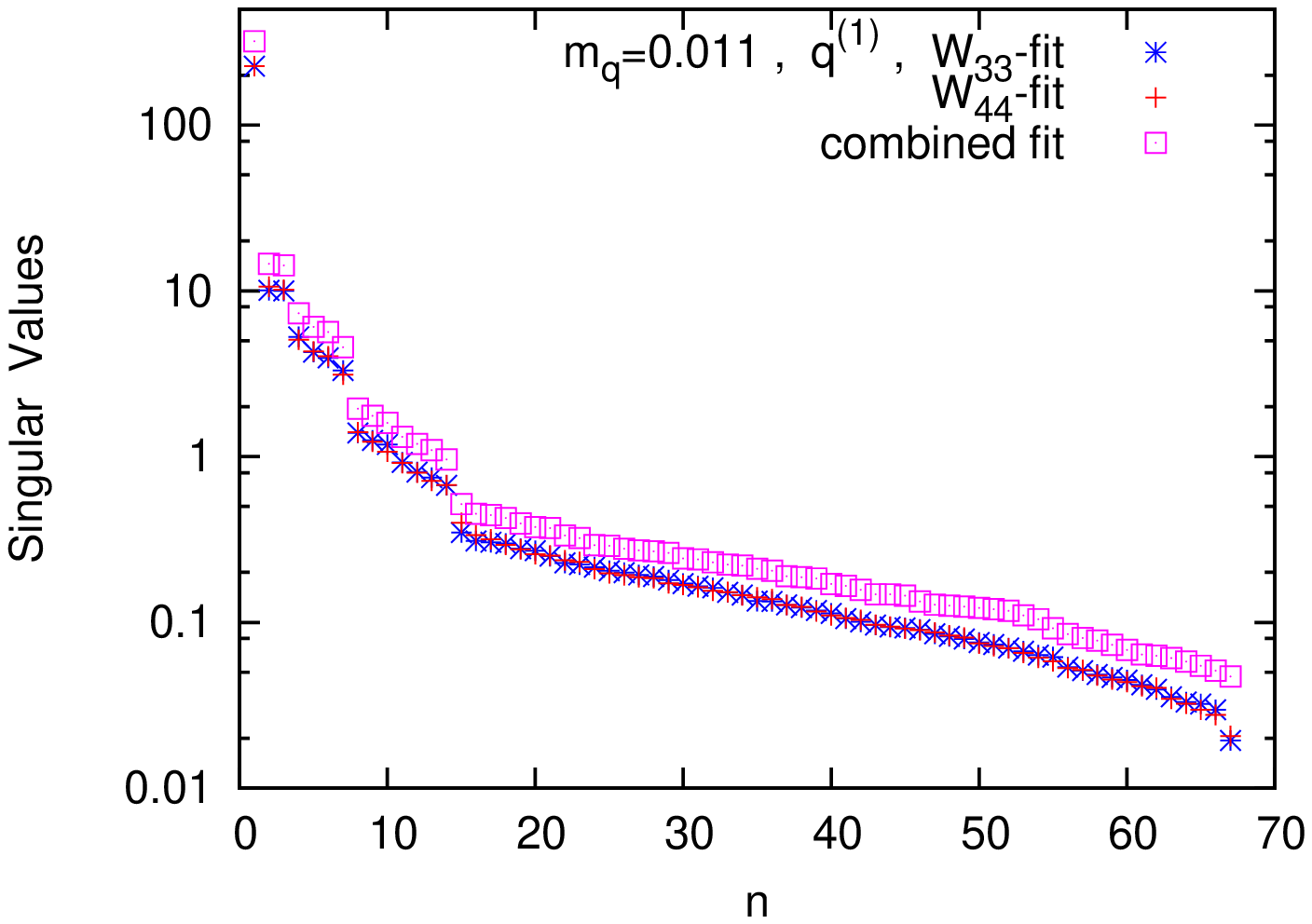}
\includegraphics[angle=270,width=.49\linewidth]{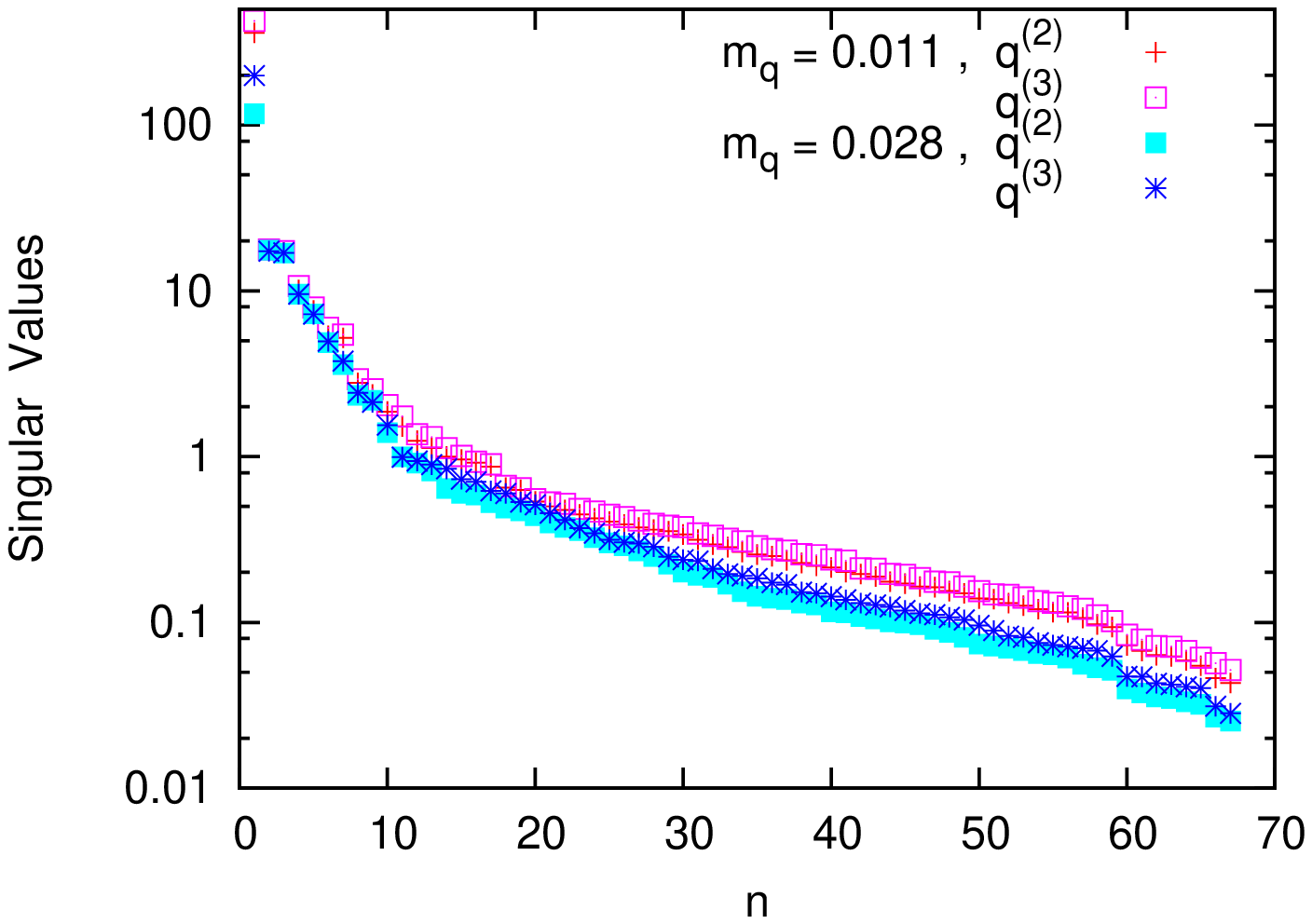}
\end{center}
\vspace*{-4mm}
\caption{Examples for the rapidly dropping magnitudes of the
SV. Due to reflection and rotation symmetries 
some Wilson coefficients coincide theoretically,
{\it e.g.}\ the coefficients of ${\cal O}_{i,33}$ in $W_{33}$,
and of ${\cal O}_{i,44}$ in $W_{44}$. We determine the
corresponding SV separately, and by a combined fit, which
implements this identity.
Left: separate and combined SV for $m_{q}=0.011$, $q^{(1)}$.
Right: combined SV for both quark masses and $q^{(2)}$,  $q^{(3)}$.}
\label{SVfig}
\end{figure}

Since \ $16 M \gg 67$ \ in all our cases, the system is strongly
over-determined. Hence we apply a Singular Value Decomposition:
it selects the $n \leq 67$ conditions with ``maximal impact''
on the solution $C_{1} \dots C_{67}$. We order the corresponding 
Singular Values (SV, analogues to eigenvalues) hierarchically. 
If their magnitude drops rapidly one has favourable conditions 
to extract a reliable result. Fig.\ \ref{SVfig} illustrates 
that this is in fact the behaviour that we observed. 
\vspace*{-2mm}

\begin{figure}[h!]
\vspace*{-4mm}
\begin{center}
\includegraphics[angle=270,width=.49\linewidth]{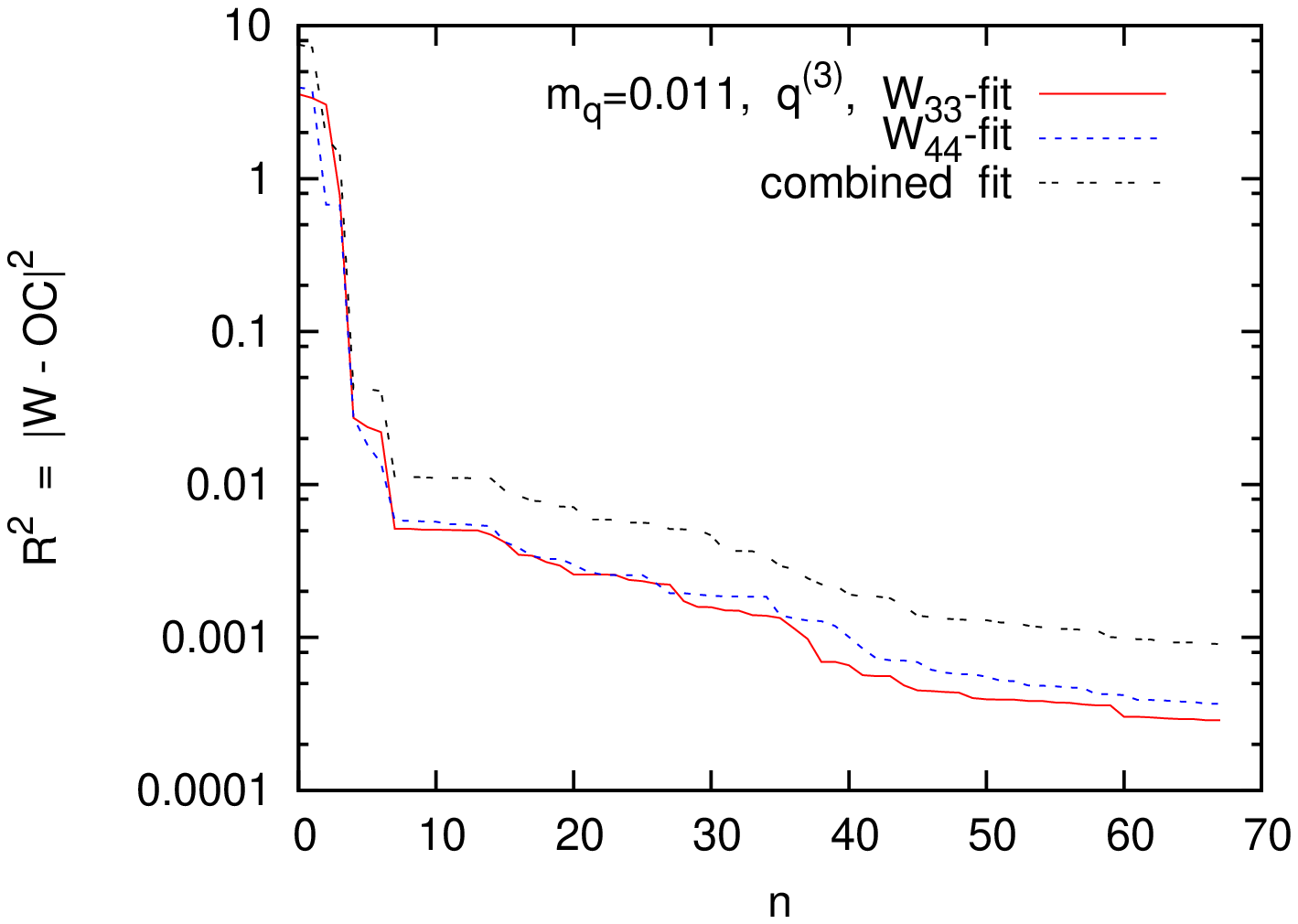}
\includegraphics[angle=270,width=.49\linewidth]{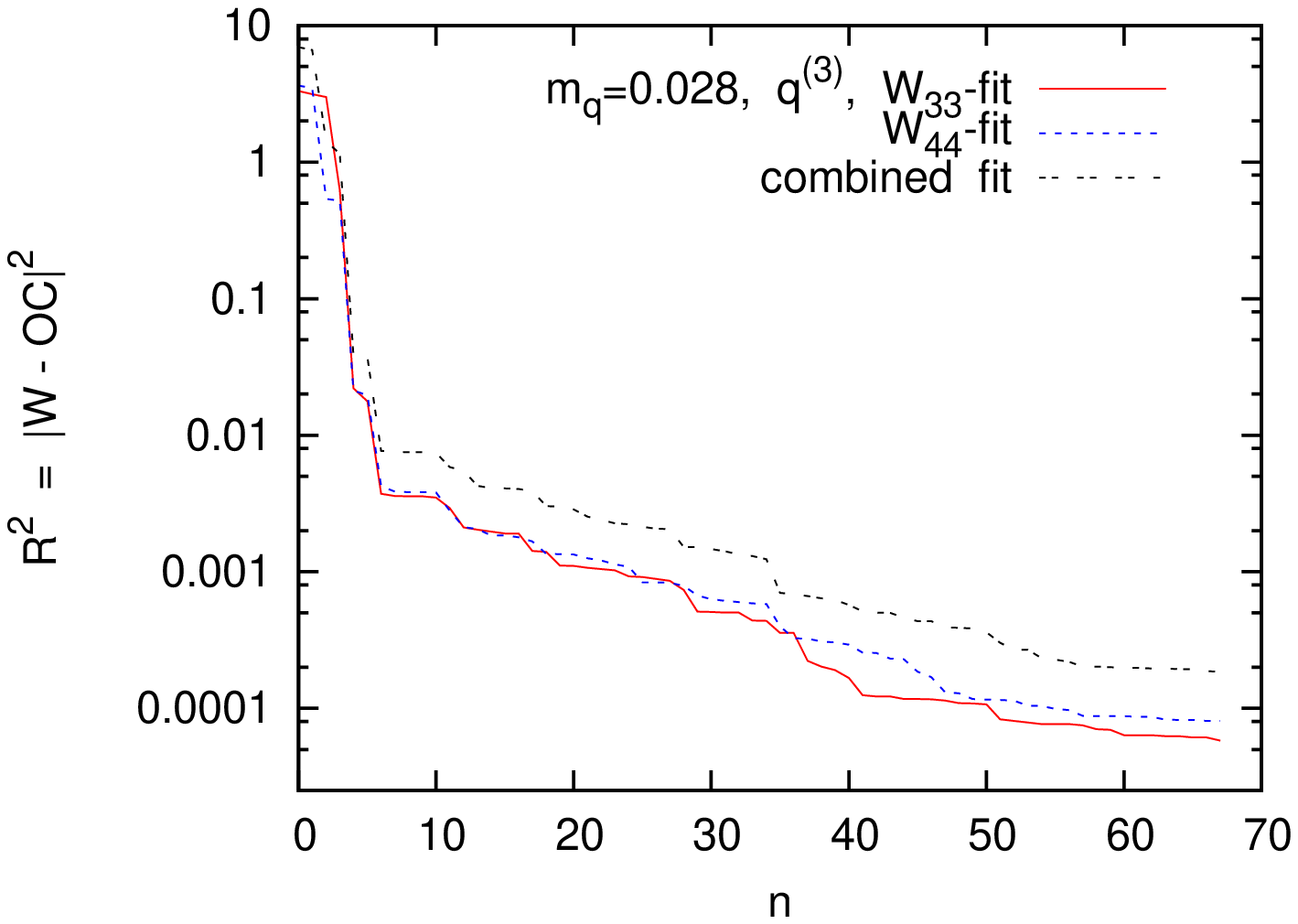}
\includegraphics[angle=270,width=.49\linewidth]{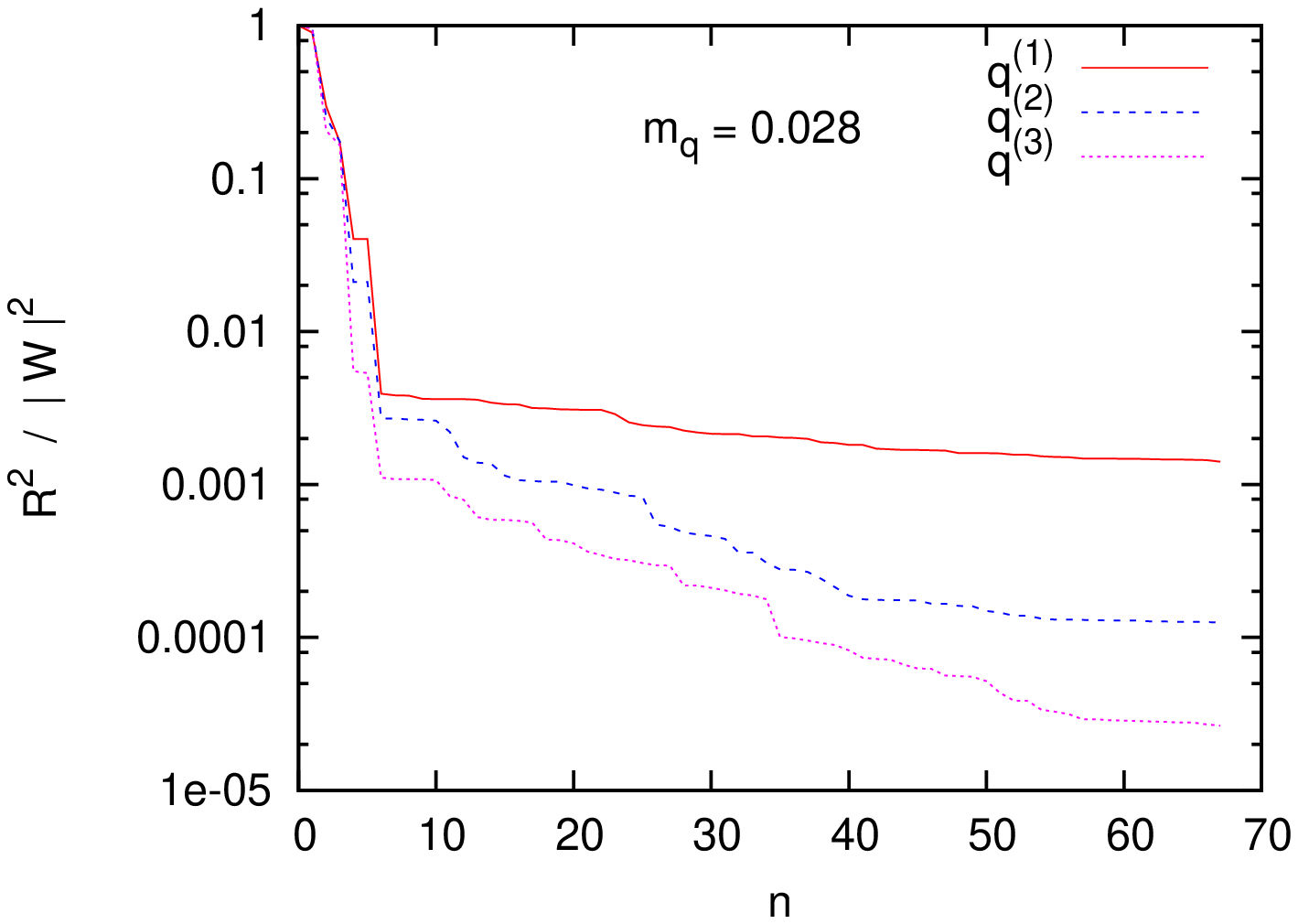}
\includegraphics[angle=270,width=.49\linewidth]{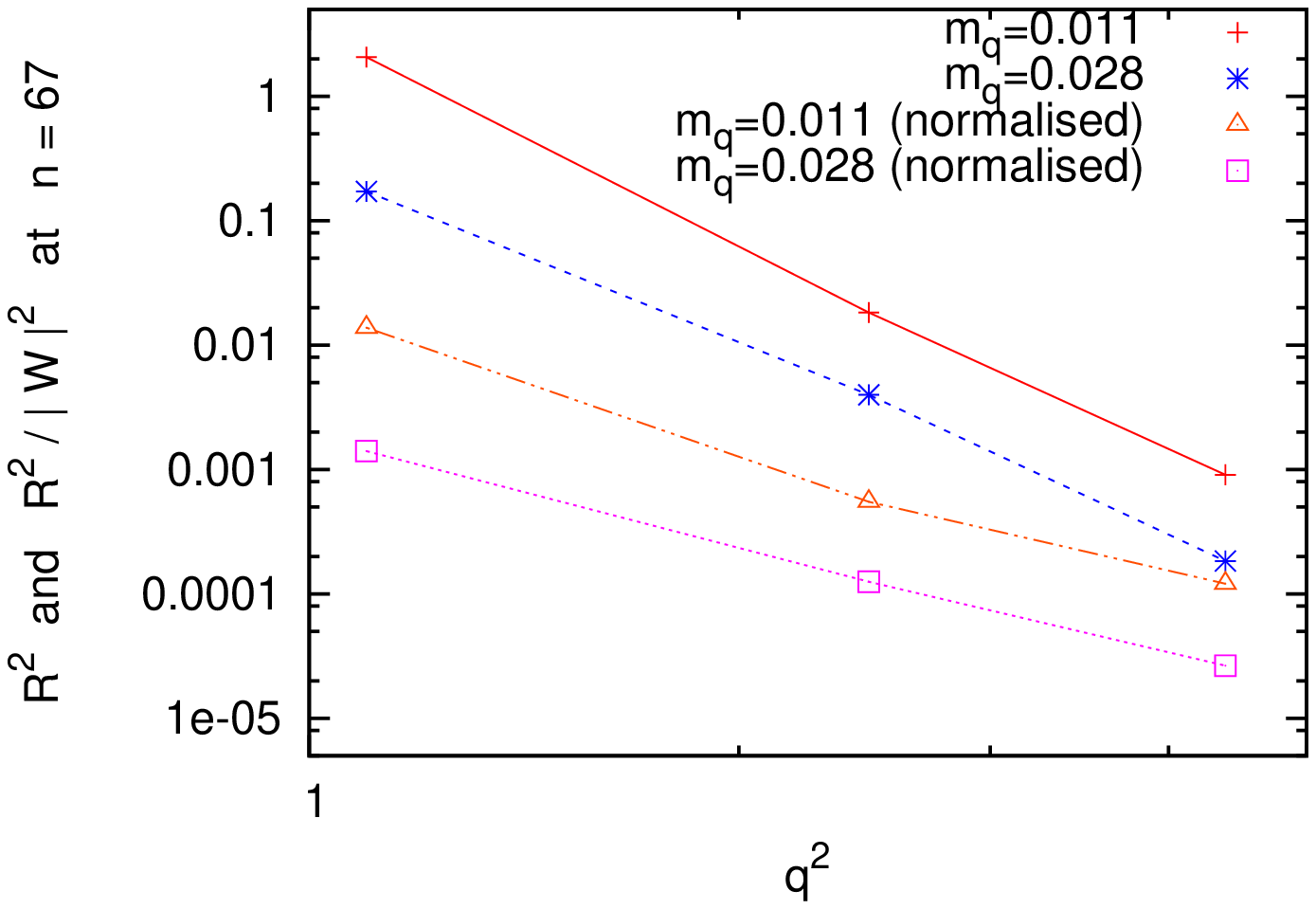}
\end{center}
\vspace*{-4mm}
\caption{Squared residues $R^{2}$ for the operator 
$\bar \psi \vec \gamma \vec D \psi$ (as an example) against 
the number $n$ of conditions (corresponding to the SV). The plots above
show results of separate and combined fits (as in Fig.\ 1), for
$q^{(3)}$ and $m_{q}=0.011$ (left), and $m_{q}=0.028$ (right). 
The plot below on the left shows the normalised $R^{2}$
for combined fits at $m=0.028$.
The final plateau value (at $n=67$) is compared
in the plot below on the right: its decrease for increasing $q^{2}$
is roughly linear, so that the relative error remains approximately 
constant.}
\label{resifig}
\vspace*{-3mm}
\end{figure}

As our next criterion, Fig.\ \ref{resifig} shows 
how the squared residues in eq.\ (\ref{fixC}) decrease as the number 
of SV involved rises from $n=1 \dots 67$.

\vspace*{-2mm}
\section{Results for the Wilson coefficients}
\vspace*{-1mm}

\begin{figure}[h!]
\vspace*{-1mm}
\begin{center}
\includegraphics[angle=270,width=.49\linewidth]{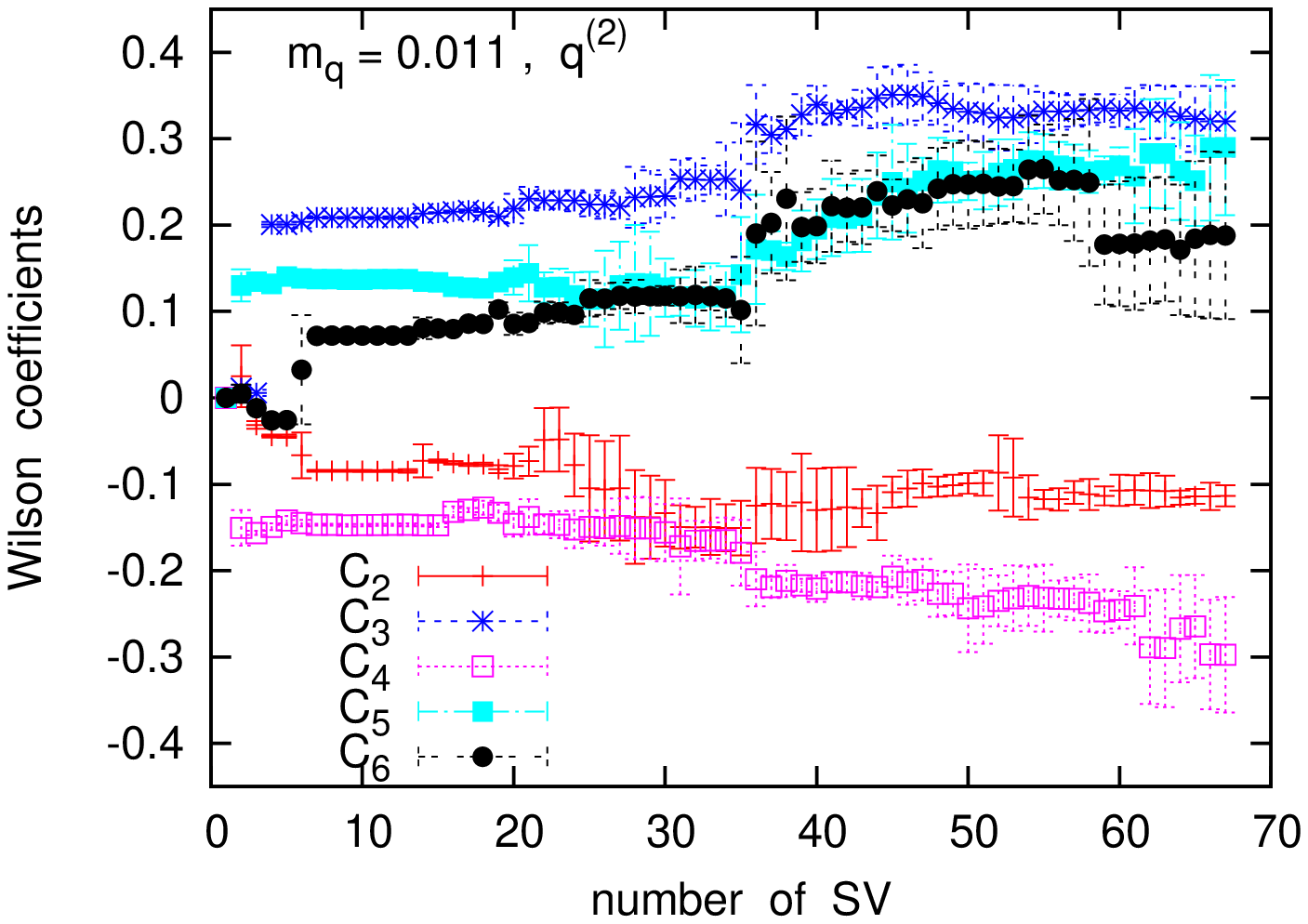}
\includegraphics[angle=270,width=.49\linewidth]{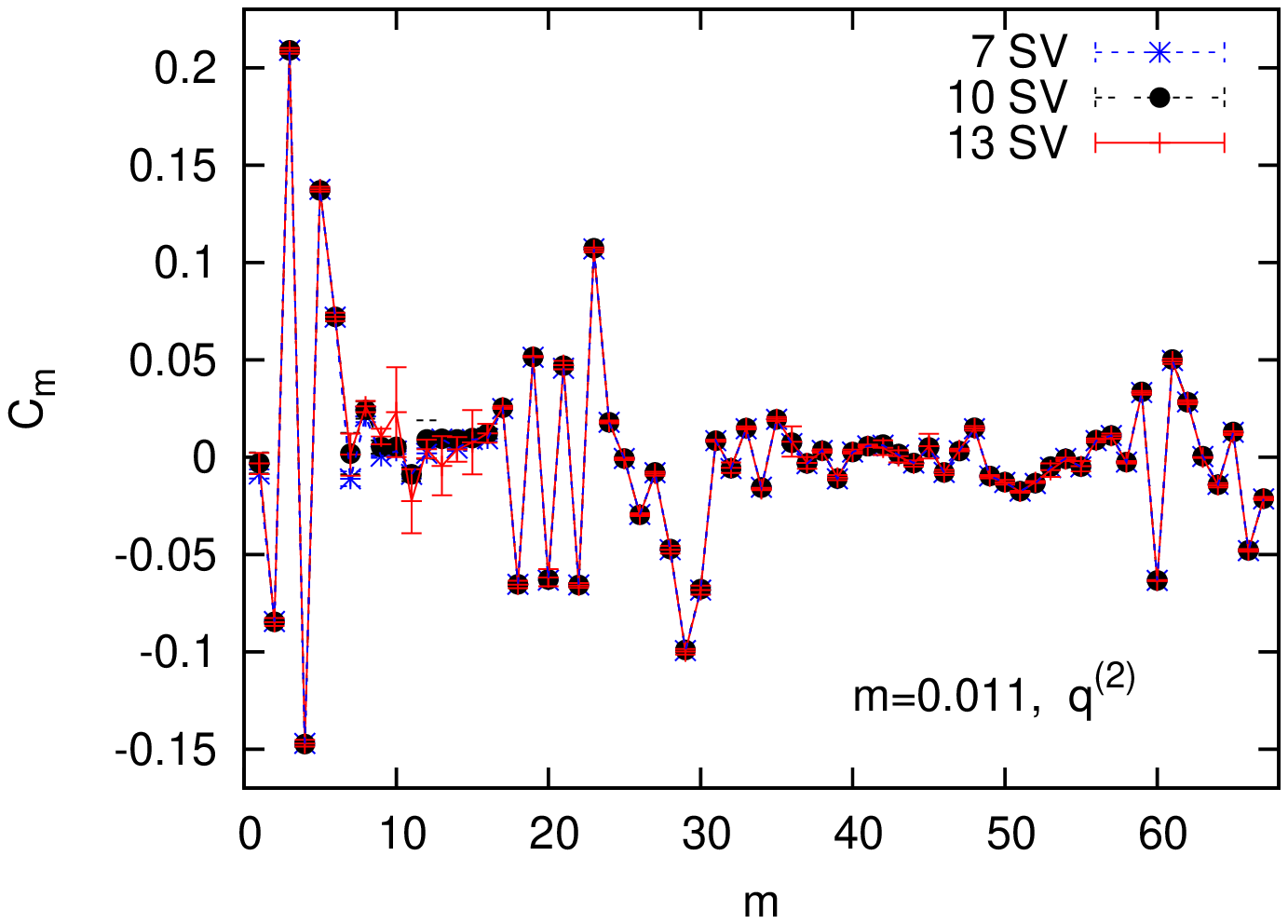}
\end{center}
\vspace*{-3mm}
\caption{Results for the Wilson coefficients at $m_{q}=0.011$ and 
$q^{(2)}$ as a varying number of SV is included. 
Left: the coefficients to the 1-derivative operators for
$n=1 \dots 67$ reveal a single plateau for $n = 7 \dots 13$.
Right: the full set $C_{1} \dots C_{67}$ agrees very well
for $n=7$, $10$ and $13$, confirming this plateau generally.}
\label{WcoefSV}
\vspace*{-1mm}
\end{figure}

Solving eq.\ (\ref{fixC}) for $C_{1} \dots C_{67}$ employs the 
inverse SV, so including all of them is not optimal: 
the tiny SV, with large relative noise, tend to distort the result. 
Therefore we computed the Wilson coefficients
with a gradually increasing number of SV, $n = 1 \dots 67$;
an example is shown in Fig.\ \ref{WcoefSV} (left). It displays the 
most important coefficients, {\it i.e.}\ those of operators with 
{\em one} derivative. The only common plateau occurs in the range of
$7 \dots 13$ SV included. To check if this plateau holds for 
all $67$ coefficients, we compare the full set obtained 
with $7$, $10$ and $13$ SV in Fig. \ref{WcoefSV} (right). 
We observe a striking confirmation 
of this plateau. 
The results look similar for other $m_{q}$ and $q$.
The optimal number of SV in each case,
considering also the impact on ${\cal M}$  
\cite{Thomas}, is displayed in Table \ref{tab}.

\begin{figure}[h!]
\begin{center}
\includegraphics[angle=270,width=.49\linewidth]{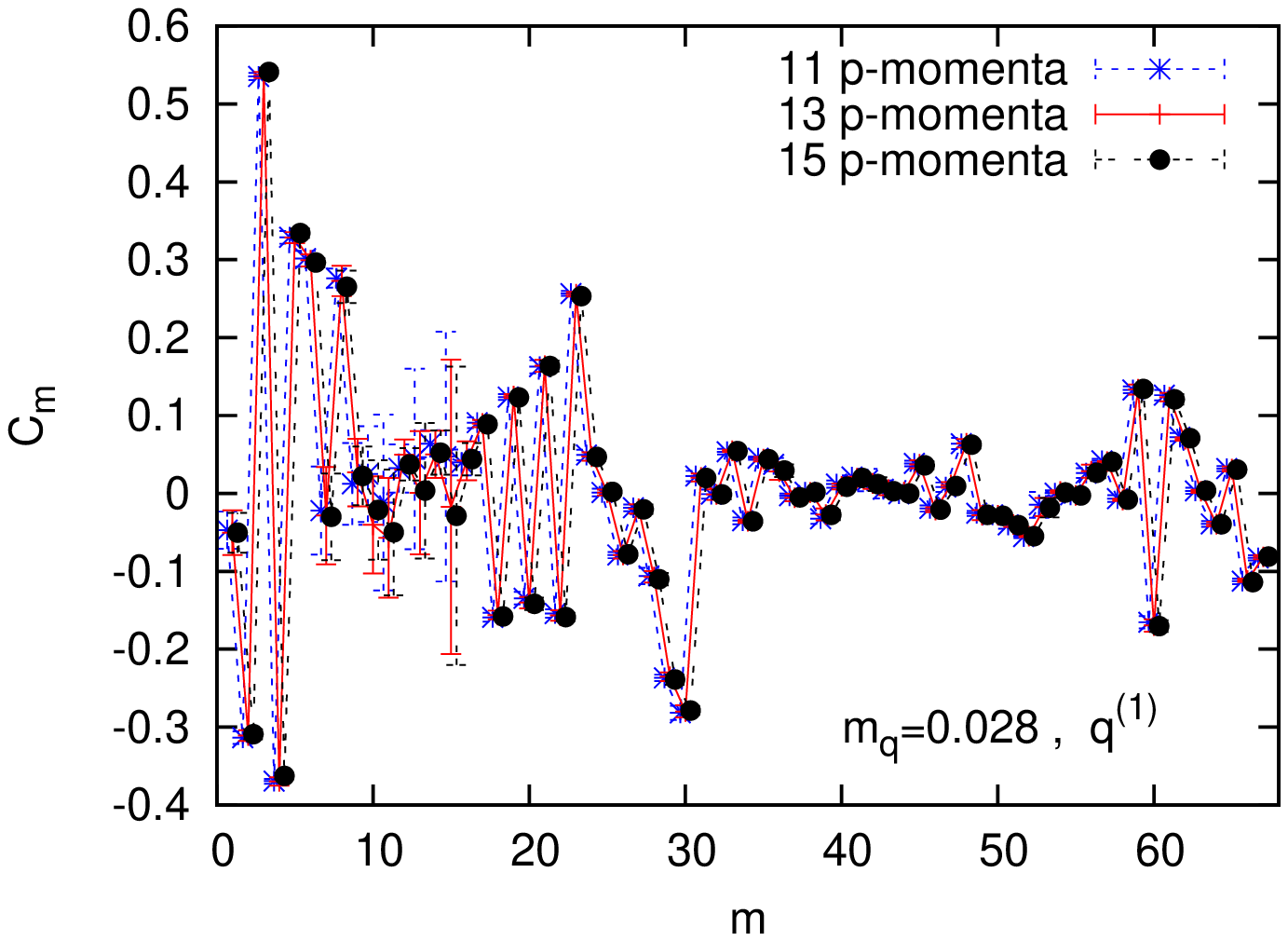}
\includegraphics[angle=270,width=.49\linewidth]{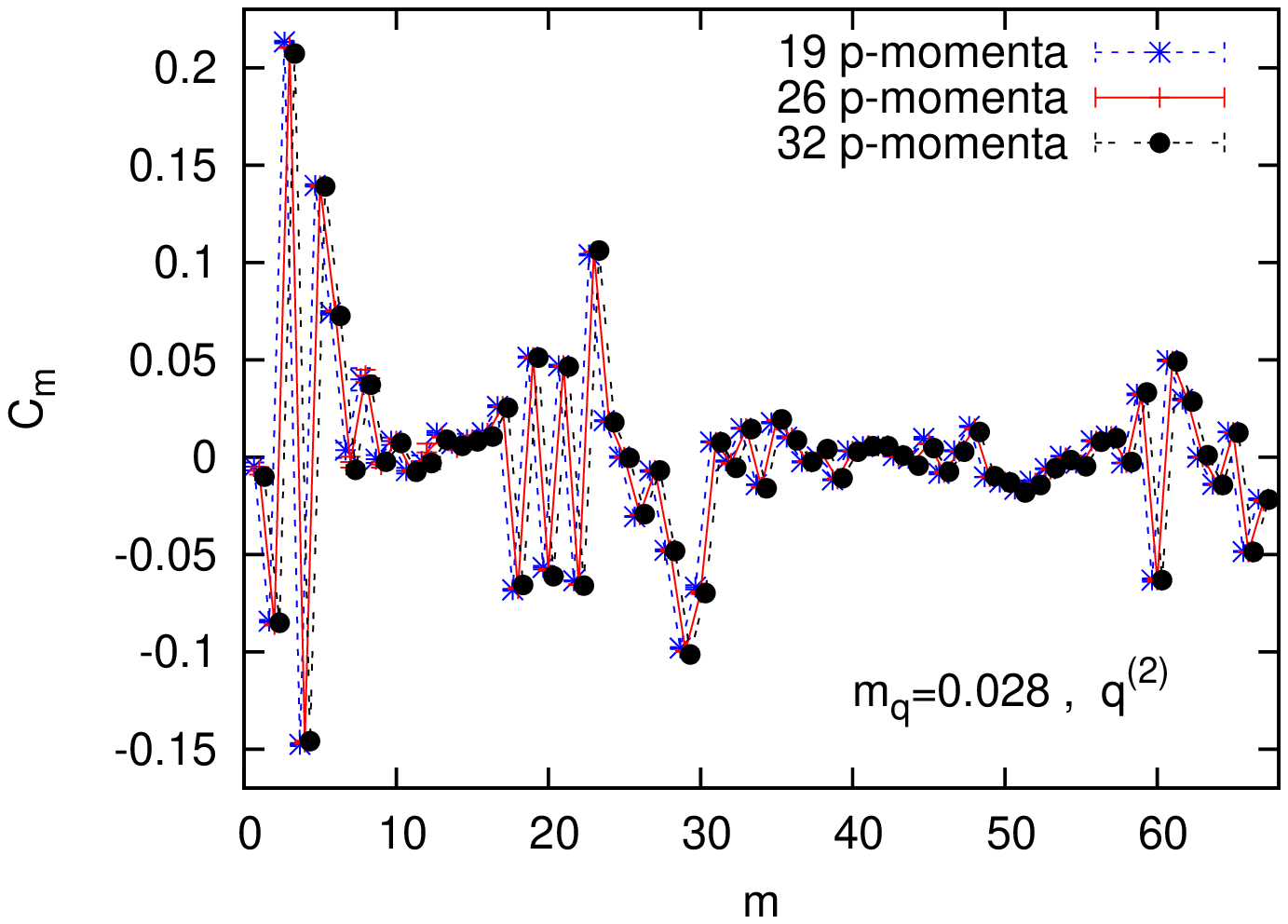}
\end{center}
\vspace*{-3mm}
\caption{The Wilson coefficients determined from
a varying number of the quark momenta.
We show results for $m_{q}=0.028$, at $q^{(1)}$ with $11$, $13$ 
or $15$ $p$-momenta (left) and at $q^{(2)}$ with 
$19$, $26$ or $32$ $p$-momenta (right). This hardly changes the
results for the Wilson coefficients, hence their convergence appears safe.}
\label{Wcoefp}
\vspace*{-4mm}
\end{figure}

Next we verify if our number of quark momenta is sufficient:
we check if the results change significantly as we omit part of them. 
Fig.\ \ref{Wcoefp} shows (with examples)
that this is not the case here: convergence for an 
increasing number of $p$-momenta is well confirmed. 

\begin{figure}[h!]
\vspace*{-2mm}
\begin{center}
\includegraphics[angle=270,width=.49\linewidth]{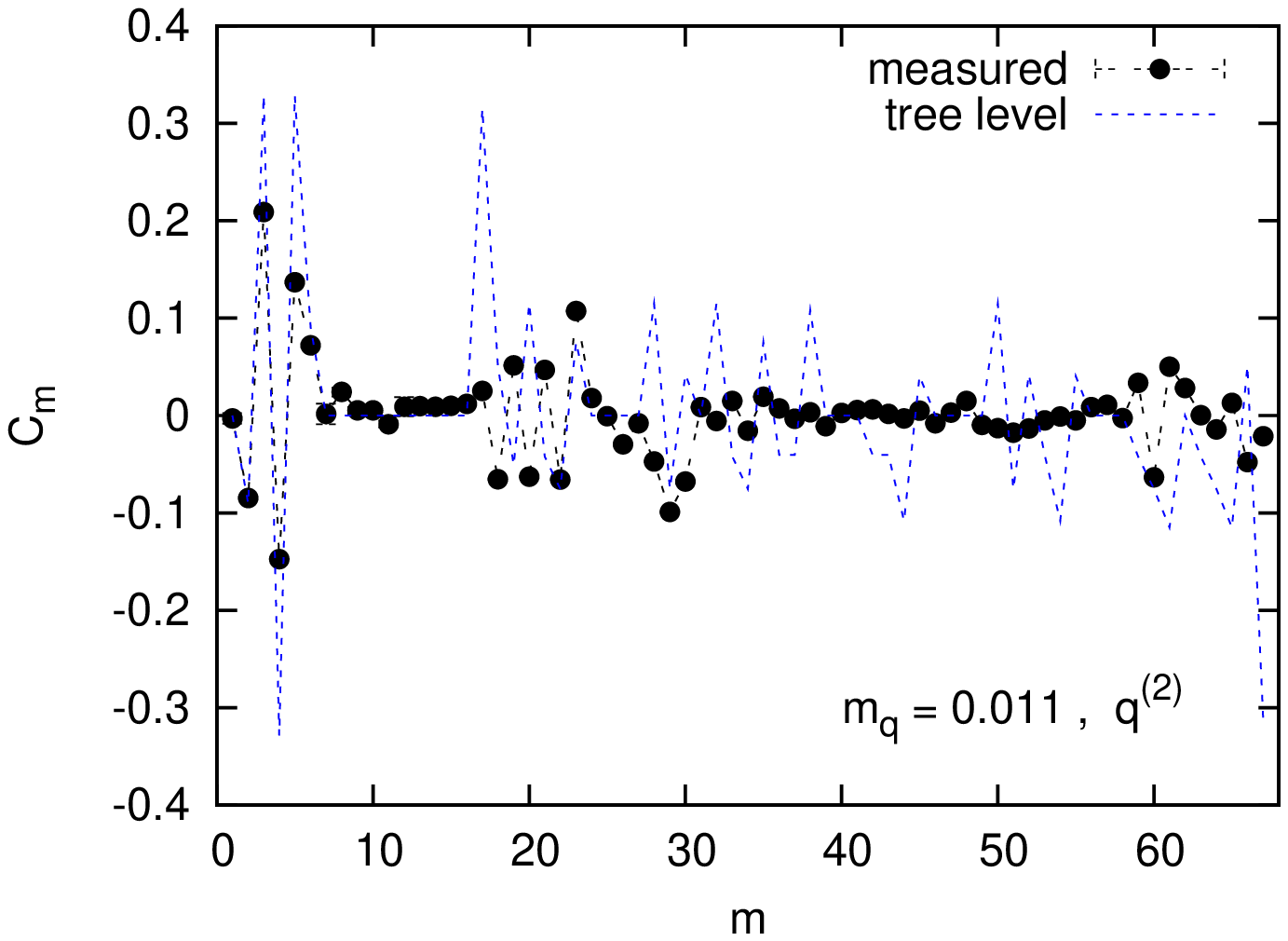}
\includegraphics[angle=270,width=.49\linewidth]{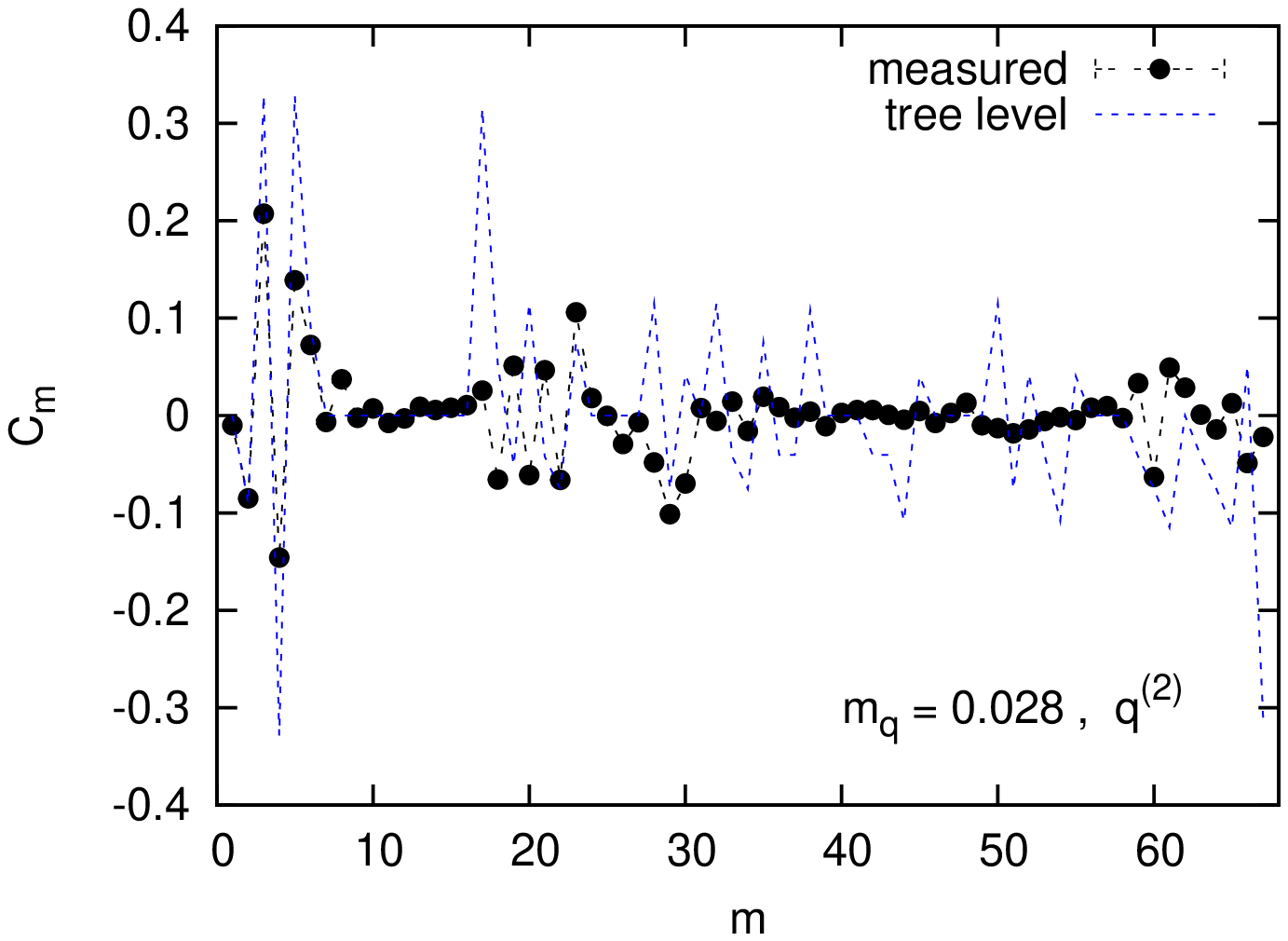}
\includegraphics[angle=270,width=.338\linewidth]{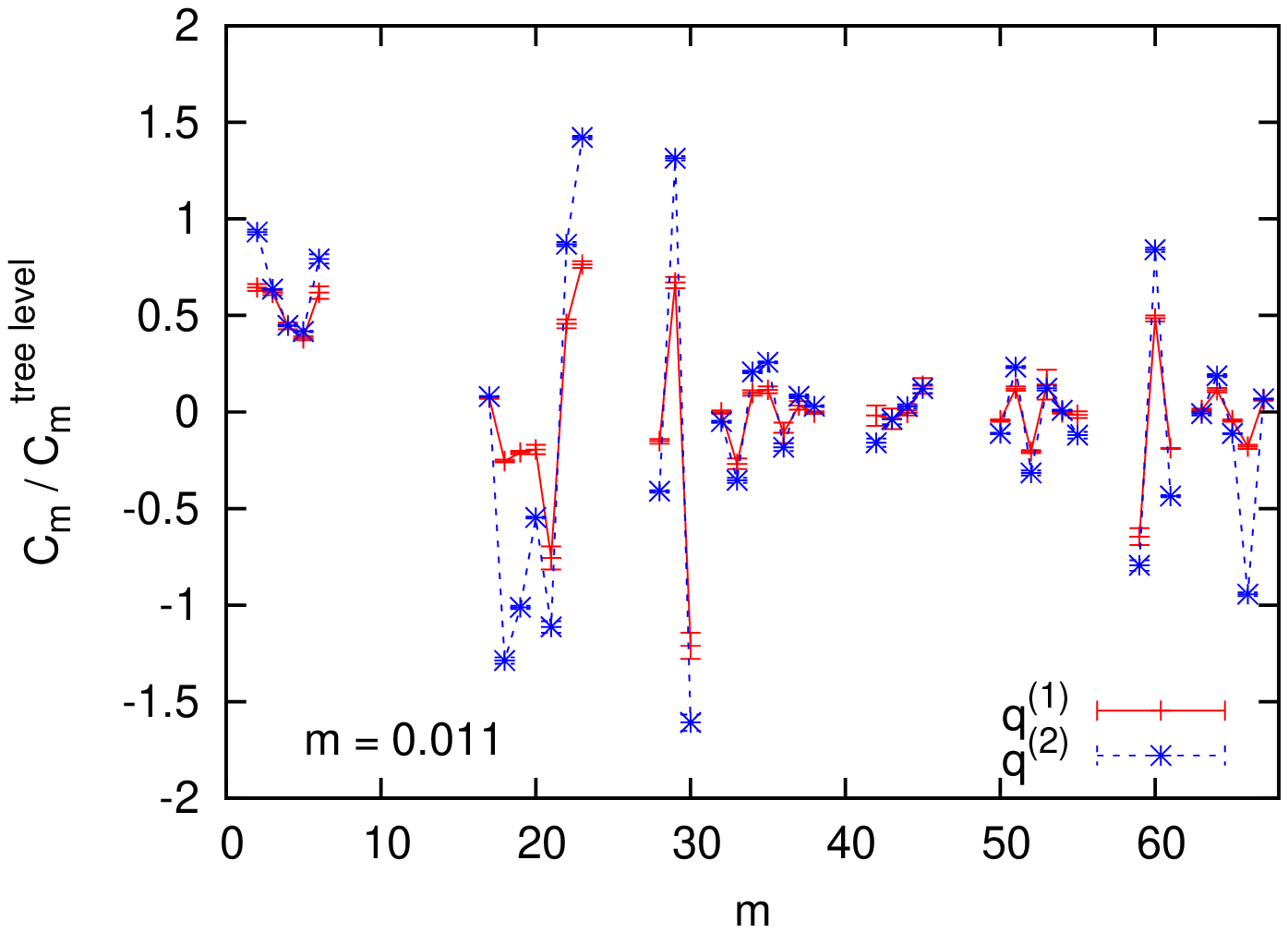}
\hspace*{-3mm}
\includegraphics[angle=270,width=.338\linewidth]{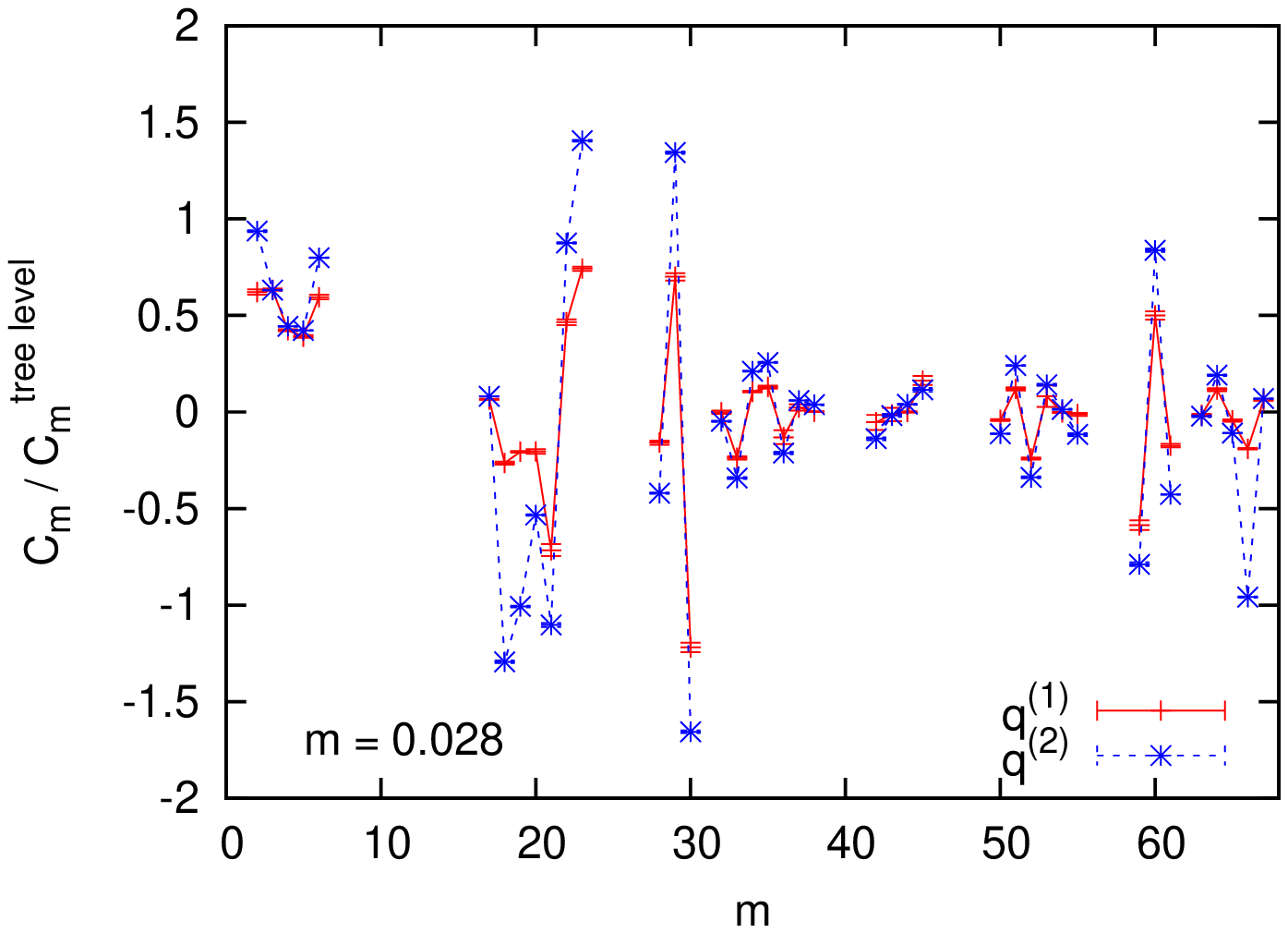}
\hspace*{-3mm}
\includegraphics[angle=270,width=.338\linewidth]{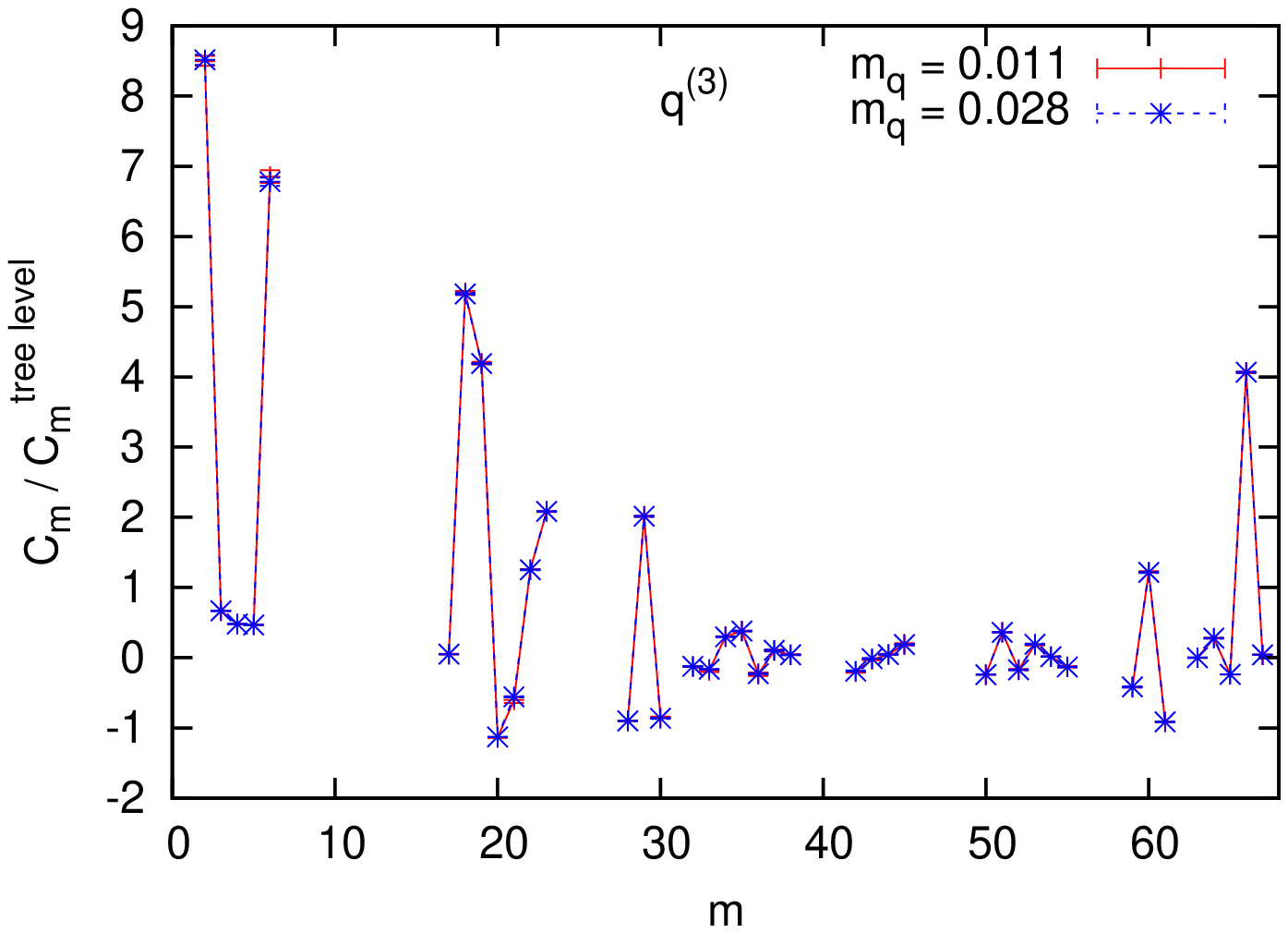}
\end{center}
\vspace*{-4mm}
\caption{Above: measured Wilson coefficients compared to their
tree level values for $q^{(2)}$. 
Coefficients for operators with an even number of
derivatives take consistently small values. Below: the ratios 
$C_{m} /C_{m}^{\rm tree~level}$ (for non-vanishing denominators).
They depend only mildly on the quark mass.}
\label{Wcoefvstree}
\vspace*{-3mm}
\end{figure}

In Fig.\ \ref{Wcoefvstree} (plots above) we compare our results for 
the Wilson coefficients at $q^{(2)}$ to the corresponding values 
at tree level. 
As we mentioned in Section 4, $C_{1}$ (which multiplies 
$\bar \psi 1 \!\! 1 \psi$) and $C_{7} \dots C_{16}$ (attached
to operators ${\cal O}^{(m)}$ with $\gamma_{\mu} \gamma_{\nu}$)
vanish at tree level in the chiral limit.
We measured consistently small values for
these coefficients, which indicates that approximate chirality
and operator mixing are indeed under control, in contrast to
previous studies with Wilson fermions \cite{Wilfer}.
Generally the measured Wilson coefficients follow the same pattern
as their counterparts on tree level, though with significantly
reduced absolute values.

Fig.\ \ref{Wcoefvstree} (plots below) show the 
commonly applied Wilson coefficients ratios
$C_{m} / C_{m}^{\rm tree~level}$. 
The dependence on the quark mass is weak, which
approves again a safe approximate chirality.
On the other hand, we observe a strong dependence on the
photon momentum $q$, as expected. Ref.\ \cite{Thomas} discusses
the detailed comparison with the theoretically expected
Bjorken scaling behaviour (cf.\ Section 4).

\vspace*{-2mm}
\section{Conclusions and outlook}
\vspace*{-1mm}

We have evaluated a set of Wilson coefficients non-perturbatively
(partial result were anticipated in Refs. \cite{Lat0708}).
They refer to twist 2; for the photon momenta that we used,
contributions by higher twists are 
suppressed \cite{Wilfer}.
The application of chiral lattice quarks has been crucial to control
the operator mixing. We demonstrated in detail that our results
are reliable regarding the number of SV and quark momenta
included in the evaluation. The measured Wilson coefficients follow 
the pattern of their counterparts at tree level, though with 
smaller absolute values.

The structure function ${\cal M}$ (in eq.\ (\ref{nustrufu}))
is now obtained by means of Nachtmann integration over $W_{\mu \nu}$
(cf.\ eq.\ (\ref{OPEeq})).
This is worked out 
for a single quark in Ref.\ \cite{Paul}. The final step to a fully
non-perturbative moment of the nucleon structure function ---
given by products between the matrix elements \cite{Lat05} and
the Wilson coefficients presented here --- 
is carried out in Refs.\ \cite{Thomas,prep}. 

\noindent
{\bf Acknowledgement :} {\it The computations for this project 
were performed on the clusters of the ``Norddeutscher
Verbund f\"{u}r Hoch- und H\"{o}chstleistungsrechnen'' (HLRN).}

\end{document}